\documentclass[12pt]{iopart}
\usepackage{graphicx}

\newcommand{\kol}{\mbox{\scriptsize K}}
\newcommand{\what}[1]{\widehat{#1}}

\newcommand{\ol}[1]{\overline{#1}}
\newcommand{\ul}[1]{\underline{#1}}
\newcommand{\vx}{\mbox{\boldmath{$x$}}}
\newcommand{\vk}{\mbox{\boldmath{$k$}}}
\newcommand{\vs}{\mbox{\boldmath{$s$}}}

\newcommand{\vxhat}{\widehat{\mbox{\boldmath{$x$}}}}

\newcommand{\yueni}{.\raisebox{1ex}{.}.}

\newcommand{\kinji}{\hspace{2pt}=\hspace{-11pt}\raisebox{1.3ex}{.}\hspace{0.5pt}\raisebox{-0.3ex}{.}\hspace{4pt}}
\newcommand{\lesssim}{<\hspace{-13pt}\raisebox{-1.2ex}{$\sim$}}
\newcommand{\vu}{\mbox{\boldmath{$u$}}}
\newcommand{\vq}{\mbox{\boldmath{$q$}}}

\newcommand{\mvec}[1]{\mbox{\boldmath{$#1$}}}
\newcommand{\bns}{\mbox{\bf NS}}
\newcommand{\rp}{\mbox{R.P.}}
\begin{document}
\title[]{The Kolmogorov turbulence theory in the light of six-dimensional Navier-Stokes' equation}
\author{
  Shunichi Tsug\'e\footnote[3]{present address : 790-3 Tohigashi, Tsukuba
    300-26 Japan, e-mail: tsuge@nyc.odn.ne.jp}
  }
\begin{abstract}
The classical turbulence theory by Kolmogorov is reconsidered using
 Navier-Stokes' equation generalized to 6D physical-plus-eddy
 space. Strong pseudo-singularity is shown to reveal itself along the
 boundary `ridge' line separating the dissipation and inertial
 sub-ranges surrounding the origin of the eddy space. A speculation is
 made that this singularity is generated by two dipoles of opposite sign
 aligned on the common axis. It is supported by the observation that the
 universal power spectrum calculated rediscovers the Kolmogorov's -5/3 power law as independent of the dimensional approach.
\end{abstract}
\section{Introduction}
As early as in 1941 Kolmogorov\cite{c1} predicted some universal
features of fluid turbulence which have been confirmed by experiments in
later years. It is rather surprising that they are derived from
dimensional analysis based on the simple assumption of `local
homogeneity' for small-scale turbulence. More surprising is the fact
that those have little to do with the equation of dynamics of fluid.

It is addressed in this paper to locate this `missing link'
through rederivation of the -5/3
 power law of the spectrum for the inertial subrange as the universal law using the dynamical equation proposed in ref.2. The key issue for this task to work is that equation on which to describe the features of small eddies as independent  of individual flow geometry.  To meet this purpose a six-dimensional Navier-Stokes equation is employed, where additional 3D space having a length dimension , corresponding to eddy size, is introduced. 

Originally such an equation has been derived elsewhere\cite{c2}, using non-equilibrium statistical mechanics starting from Liouville's equation. In this formalism the 3D `eddy' space has been introduced as natural consequence of a mathematical procedure, namely, the separation of variables of turbulent fluctuation-correlation equation. The equation stands as 6D generalization of the Navier-Stokes equation which in 3D physical space degenerates to the classical  equation.

In what follows, however, it is intended to rederive the same equation using phenomenologies alone on the basis same as the classical theory of K\'arm\'an and Howarth\cite{c3}. Then, the wave-number space is introduced for separating variables (Sec.2). It is then Fourier-transformed into eddy space, thereby the 6D Navier-Stokes equation, the basis to all what follows  is established (Sec.3). The equation gives a novel expression for turbulent dissipation, enabling to predict existence of a pseudo-singularity surrounding the dissipation region (Sec.4). In inertial subrange, this singularity turns out to be a couple of dipoles of opposite sense, separated by the order of several tens to hundreds the Kolmogorov length (Sec.5). Power spectrum of this locally homogeneous turbulence is calculated, showing a wave-number dependence close to -5/3 power law of the Kolmogorov theory.(Sec.6)  
\section{K\'arm\'an-Howarth formalism revisited as governing inhomogeneous turbulence}
In 1938, von K\'arm\'an and Howarth proposed an equation governing
homogeneous isotropic turbulence whose original form is written
as\cite{c3}
\begin{equation}
 \ol{u_j'(\what{\mbox{NS}})_l + \hat{u}_l'(\mbox{NS})_j}=0
\end{equation}
In this equation $u_j'=u_j'(\vx,t)$ and $\hat{u}_l'=u_l'(\vxhat,t)$ are
instantaneous velocity fluctuation at $\vx$ and $\vxhat$ , respectively,
overbar ($\rule[2.3mm]{5mm}{.1mm}$) denotes the conventional (ensemble) average and     
\begin{eqnarray}
\bns &\equiv& \bns(\nabla , \ul{\vu}, \ul{p})\nonumber\\
 &=&(\frac{\partial}{\partial t}+\ul{\vu}\cdot\nabla -\nu\nabla^2)\ul{\vu} + \rho^{-1}\nabla \ul{p}=0
\end{eqnarray}
denotes the Navier-Stokes equation written in terms of the instantaneous fluid quantities
\begin{equation}
\begin{array}{rcl}
 \ul{\vu}&=&\vu+\vu'\\
 \ul{p}&=&p+p'
\end{array}
\end{equation}
where $\vu$ and $p$  are the average velocity and pressure,
respectively, $\nu$  is the kinematic viscosity, and $\nabla$ denotes
nabla vector. Likewise $\what{\bns}$ is defined by Eq.(2) in
which ( $\nabla,\ul{\vu},\ul{p}$ ) is replaced with (
$\hat{\nabla},\hat{\ul{\vu}}, \hat{\ul{p}}$).

Eq.(1) is an equation in 6-D physical space and time, namely, has seven
independent variables ($\vx, \hat{\vx}, t$).
The classical theory\cite{c3} chose to discuss homogeneous and
isotropic turbulence to 
reduce independent variables to ($|\vx-\hat{\vx}|,t$). We can
show, however, 
that the equation can be renovated to be available for
inhomogeneous turbulence as well by introducing separation of
variables, together with a closure condition which seems
plausible intuitively.

The actual derivation proceeds as follows: Eq.(1) in which
$\bns(\nabla, \ul{\vu}, \ul{p})$ is decomposed into average and fluctuating parts using (3),reads 
\begin{equation}
 \ol{u_j'(\what{\mbox{NS}})_l' + \hat{u}_l'(\mbox{NS})_j'} = 0
\end{equation}
	where
\begin{equation}
 (\mbox{NS})_j'\equiv(\frac{\partial}{\partial t}
  +u_r\frac{\partial}{\partial x_r}
  -\nu\frac{\partial^2}{\partial x_r^2})u_j'
  +\frac{1}{\rho}\frac{\partial p'}{\partial x_j}
  +\frac{\partial}{\partial x_r}(u_j'u_r')
\end{equation}
together with $(\what{\mbox{NS}})_l'$   defined as in Eq.(1). Upon
substituting Eq.(5) into Eq.(4) we see that the equation consists of
terms of double and triple fluctuation correlations, 
 which we will decompose as
\begin{eqnarray}
 \ol{u_j'\hat{u}_\lambda'}&=&\rp l^6
  \int_{-\infty}^{\infty}
  g_j(\vx,\vk)g_\lambda(\hat{\vx},\hat{\vk})\delta(\vk+\hat{\vk})
  d\vk d\hat{\vk}\;,\;\lambda=(l,4)\\
 \ol{u_j'\hat{u}_l'\tilde{u}_r'}&=&\rp l^9
  \int_{-\infty}^{\infty}
  g_j(\vx,\vk)
  g_l(\hat{\vx},\hat{\vk})
  g_r(\tilde{\vx},\tilde{\vk})
  \delta(\vk+\hat{\vk}+\tilde{\vk})
  d\vk d\hat{\vk}d\tilde{\vk}
\end{eqnarray}
where $\vk$ is the wave number, $l$ is the characteristic length of flow
geometry, R.P. denotes taking real part, and subscript $\lambda$ runs 1
through 4. A supplementary definition
\begin{equation}
 u_4'=\rho^{-1}p'
\end{equation}
stands for the pressure fluctuation.
It is more convenient for later use to employ the following alternative
expressions:
\[
 \ol{u_j'\hat{u}_\lambda'}=\rp l^3
  \int_{-\infty}^{\infty}
  g_j(\vx,\vk)g_\lambda(\hat{\vx},-\vk)d\vk\;,\;\lambda=(l,4)
\hspace{35mm}(6')
\]
\[
\hspace*{-13mm} \ol{u_j'\hat{u}_l'\tilde{u}_r'}=\rp l^6
  \int_{-\infty}^{\infty}
  g_j(\vx,\vk)d\vk
  \int_{-\infty}^{\infty}
  g_l(\hat{\vx},-\vk+\hat{\vk})
  g_r(\tilde{\vx},-\hat{\vk})d\hat{\vk}\;,
  \;(\tilde{\vx};\vx \;\mbox{or}\; \hat{\vx})
  \hspace{4.5mm}(7')
\]

It is to be noted that there is no a priori reason why the double and
triple correlations are to be expressed in terms of the {\it same} $g_\lambda$. At this point it is simply invoked as the closure condition to
truncate the chain of equations at the level of Eq.(4). This closure can
be justified only through the quality of outcome as to whether it fits
with physical reality of turbulence. 

With these preliminaries, Eq.(4) with assumption (6',7') substituted
into, leads to the form typical of separation of variables, here into
($\vx$,t) and ($\hat{\vx}$ ,t), respectively; 
\begin{equation}
 \int_{-\infty}^{\infty}
d\vk g_j\hat{g}_l
[\underbrace{g_j^{-1}(\mbox{ns})_j^{(0)}}_{\displaystyle =\;i\omega}+
 \underbrace{\hat{g}_l^{-1}(\what{\mbox{ns}})_l^{(0)}}_{\displaystyle =\;-i\omega}]=0
\end{equation}
where $\omega$  is the separation parameter having the dimension of the
frequency, and $\hat{g}_l$  stands for $g_l(\hat{\vx},\vk)$ , and
$(\mbox{ns})^{(0)}$  is defined by
\begin{equation}
\hspace*{-15mm}
 (\mbox{ns})_j^{(0)}\equiv [\mbox{ns}(\mvec{g},g_4)^{(0)}]_j
  =(\frac{\partial}{\partial t}+u_r\frac{\partial}{\partial x_r}-\nu\frac{\partial^2}{\partial x_r^2})g_j
  +\frac{\partial u_j}{\partial x_r}g_r
  +\frac{\partial g_4}{\partial x_j}
  +\frac{\partial}{\partial x_r}I(g_jg_r)
\end{equation}
with
\[
 I(g_jg_r)\equiv l^3\int_{-\infty}^{\infty}
 g_j(\vk-\hat{\vk})g_r(\hat{\vk})d\hat{\vk}
\]
Imaginary factor $i$ as appearing in Eq.(9) reflects the statistical symmetry of the tensor $\ol{u_j'\hat{u}_l'}$ , which is met under the following additional condition
\begin{equation}
 g_l(\hat{\vx},-\vk)=[g_l(\hat{\vx},\vk)]^*
\end{equation}
In fact, then, the two terms inside[\hspace{3mm}] of integral (9) are
commutable to each other through taking complex conjugate(*),  if
$i\omega$ is purely imaginary.  Thus we are led to the assertion that
the only equation that need to be solved is
\begin{equation}
 i\omega g_j=(\mbox{ns})_j^{(0)}
\end{equation}

To go further
 from this point on, we need to have relationship between
frequency  $\omega$ introduced as the separation parameter and wave number
 $\vk$ by which the solution is constructed in the form of bilinear
integral. Physically they are related to each other by the dispersion
relation $\omega(\vk)$ . Or, instead, we may introduce phase velocity $\mvec{c}$ by
\begin{equation}
 \omega=\mvec{c}\cdot\vk
\end{equation}
with no loss of generality. In fact, its constancy does not mean the
turbulent eddies being assumed as nondispersive. For the
unsteady term in the equation $(\partial/\partial t \neq 0)$ is
responsible for the dispersive part, if any. 

Owing to the convolution form of integral (10) periodic factor drops off
from Eq.(12) despite its nonlinear structure by putting
\begin{equation}
 g_\lambda(\vx,\vk)=e^{i\vk\cdot\vx}f_\lambda(\vx,\vk)\;,\;\lambda=(l,4)
\end{equation}
thereby we have the following equation governing its amplitude $f_\lambda$,
\begin{equation}
 i\omega f_j=(\mbox{ns})_j
\end{equation}
where
\begin{eqnarray}
 (\mbox{ns})_j &\equiv&[\mbox{ns}(\mvec{f},f_4)]_j\nonumber\\
 &=&[\frac{\partial}{\partial t}
  +u_r\partial_r(\vk)-\nu\partial_r^2(\vk)]f_j
  +\frac{\partial u_j}{\partial x_r}f_r
  +\partial_j(\vk)f_4
  +\partial_r(\vk)I(f_jf_r)
\end{eqnarray}
with
\begin{equation}
 \mvec{\partial}(\vk)\equiv\nabla(\vx)+i\vk
\end{equation}
It is readily checked that expressions defined by (15) and (10) are related to each other by 
\[
 (\mbox{ns})_j^0
 =[(\mbox{ns})_j]_{\footnotesize{\mvec{\partial}(\vk)\rightarrow \nabla(\vx)}}
\]

Several remarks are in order with regards to physical implications of
variables having appeared in this section: Amplitude function $\mvec{f}$  as
obeying Eq.(14) is {\it essentially} complex and is {\it not} an
observable. It is related to the observable quantity of fluctuation
correlation through 
\begin{equation}
 \ol{u_j'\hat{u}_l'}=\rp l^3\int_{-\infty}^{\infty}
  e^{i\vk\cdot(\vx-\hat{\vx})}
  f_j(\vx,\vk)f_l^*(\hat{\vx},\vk)d\vk
\end{equation}
as is confirmed by (6'),(11)and (13). Its complex variable structure has its
origin in Eq.(9)
where the imaginary unit $i$ is introduced from the symmetry postulate
of statistical  mechanics\cite{c2} for correlation tensor (18). We may
note some coincidental parallelism to Schroedinger's wave equation where
the imaginary factor secures the corresponding tensor to be Hermitean as
it should.

The most crucial on which this paper rests is the soundness of the basis
of Eq.(1).   Originally it was a product of intuition by Theodore von
K\'arm\'an without any `first principle' ground shown in the
paper\cite{c3}. In fact, then, any linear combinations of $(\bns)'$  would be claimed as equally qualified. Firm basis is provided by nonequilibrium statistical mechanics\cite{c4} to justify that the linear combination only of the form (1) is consistent with Liouville's equation, namely, the equation of continuity in the phase space of Hamiltonian mechanics.  . 
\section{Turbulence in eddy space}
The nonlinear integro-differential equation we have derived in the previous section is not very easy to deal with. Since, however, the integral is of convolution type, it transforms to a simple product, through Fourier transform as 
\begin{equation}
 f_j(\vx,\vk)=(2\pi l)^{-3}\int_{-\infty}^{\infty}
  d\vs e^{-i\vk\cdot\vs}q_j(\vx,\vs)
\end{equation}
The actual operation of the transform on Eq.(14) gives
\begin{equation}
 [-c_r\frac{\partial}{\partial s_r}
  +\frac{\partial}{\partial t}
  +u_r\partial_r(\vs)
  -\nu \partial_r^2(\vs)]q_j
  +\partial_j(\vs)q_4
  +\frac{\partial u_j}{\partial x_r}q_r
  +\partial_r(\vs)(q_jq_r)=0
\end{equation}
where
\begin{equation}
 \partial_j(\vs)\equiv\partial/\partial x_j+\partial/\partial s_j
\end{equation}
Variable $\vs$  introduced in Eq.(19) as the Fourier variable adjoint to
wave number $\vk$   has dimension of length, which stands for eddy size
with direction of the vorticity vector, so may well be called {\it eddy
variable}. It should be remarked that Eq.(20) is the Navier-Stokes
equation in 6D (physical plus eddy) space, describing the motion of
turbulent vortices moving with phase velocity $\mvec{c}$  . In fact, this equation
governing $q_\lambda(\vx,\vs)$ for prescribed $\vu(\vx)$ and $p(\vx)$ is
alternatively written as
\begin{equation}
 [\mvec{c}\cdot\nabla(\vs)]\mvec{q}=
  \bns(\mvec{\partial}(\vs),\vu+\mvec{q},p+\rho q_4)
  -\bns(\nabla(\vx),\vu,p)
\end{equation}
with
\begin{equation}
\mvec{\partial}(\vs)\equiv\nabla(\vx)+\nabla(\vs)
\end{equation}
where $\bns$  has been defined by Eq.(2), and $\nabla$ denote the nabla
vector in the respective spaces\footnote[3]{The bold-face letters denote
vectors having three components, not to be confused with 
6D vectors.}.

Simplicity in expression for physical quantities is another advantage of
working with $\vs$-space. For example, fluctuation-correlation formula
(18) takes remarkably simple form 
\begin{equation}
 \ol{u_j'\hat{u}_l'}=(2\pi l)^{-3}\int_{-\infty}^{\infty}
  d\vs q_j(\vx,\vs+\vx)q_l(\hat{\vx},\vs+\hat{\vx})
\end{equation}
as is easily confirmed by substituting (19) and employing definition of the delta function 
$\delta(\vs/l)=(2\pi l)^3 \int e^{i\vk\cdot\vs}d\vk$ . Turbulent
dissipation which has been dealt with as an elementary parameter in the
classical dimensional analysis can be expressed in this space by an explicit form:
We have, by definition,  
\begin{equation}
 \begin{array}{rcl}
  \displaystyle
  \frac{\epsilon}{2}&= &
\displaystyle\frac{\nu}{2}
   \ol{\left(\frac{\partial u_j'}{\partial x_l}
   +\frac{\partial u_l'}{\partial x_j}\right)^2} \\
  \vspace{0mm}\\
  &=& 
\displaystyle\frac{\nu}{2}
   \ol{\left[\left(\frac{\partial u_j'}{\partial x_l}
   +\frac{\partial u_l'}{\partial x_j}\right)
   \left(\frac{\partial \hat{u}_j'}{\partial \hat{x}_l}
   +\frac{\partial \hat{u}_l'}{\partial \hat{x}_j}\right)\right]
   }_{\hat{\vx}=\vx} \\
  \vspace{0mm}\\
  &=& 
\displaystyle \nu \left[
   \left(\frac{\partial^2 \ol{u_j'\hat{u}_j'}}
    {\partial x_l\partial \hat{x}_l}\right)
   +\left(\frac{\partial^2 \ol{u_j'\hat{u}_l'}}
    {\partial x_l\partial \hat{x}_j}
   \right)\right]
   _{\hat{\vx}=\vx}
 \end{array}
\end{equation}
On the other hand, we have from (24)
\[
\begin{array}{rcl}
\displaystyle
\frac{\partial^2 \ol{u_j'\hat{u}_l'}}
    {\partial x_m\partial \hat{x}_n}
 &= &\displaystyle
 \frac{1}{(2\pi l)^3}
 \int_{-\infty}^{\infty} d\vs \left(\frac{\partial q_j}{\partial x_m}
+\frac{\partial q_j}{\partial s_m}\right)
\left(\frac{\partial \hat{q}_l}{\partial \hat{x}_n}
 +\frac{\partial \hat{q}_l}{\partial \hat{s}_n}\right)_{\hat{\vs}=\vs}\\
  \vspace{0mm}\\
 &= &\displaystyle
 \frac{1}{(2\pi l)^3}
 \int_{-\infty}^{\infty} d\vs
 \big[\partial_m (\vs)q_j\big]
 \big[\hat{\partial}_n (\hat{\vs})\hat{q}_l\big]_{\hat{\vs}=\vs}
\end{array}
\]
Thus we are led to the final expression for dissipation $\epsilon$ as
\begin{equation}
\frac{\epsilon}{2}=\frac{\nu}{(2\pi l)^3}\int_{-\infty}^{\infty}
 d\vs [(\partial_l(\vs)q_j)^2+(\partial_l(\vs)q_j)
 (\partial_j(\vs)q_l)]
\end{equation}
This formula will reveal a new facet of the dissipation function hidden in the classical working space. 
\section{Dimensional consideration on the existence of pseudo-singularity}
In the Kolmogorov regime, dissipation $\epsilon$  was introduced,
together with kinematic viscosity $\nu$  , characteristic velocity  $v$
and length $l$    , respectively, as an elementary parameter by which to
construct the dimensional analysis. They are related to each other
through the following relationships 
\begin{equation}
 \epsilon\sim v^3/l \sim v_{\kol}^3/l_{\kol}
\end{equation}
where the subscript K denotes the respective quantities in the dissipation region. The following formulae are their immediate consequences
\begin{equation}
 \left.
 \begin{array}{rcl}
  v_{\kol}l_{\kol}&= &\nu \\
  v_{\kol}&= &vR^{-1/4} \\
  l_{\kol}&= &lR^{-3/4}\hspace{10mm} \mbox{(Kolmogorov length)}
 \end{array}\right\}
\end{equation}
where $R$ is the flow Reynolds number 
\begin{equation}
 R=vl/\nu
\end{equation}

Formula (26) we have derived in the newly defined hyperspace sheds some
lights on this classical theory that is conducted within the physical
space, in the sense that $\epsilon$  is not necessarily an elementary
parameter any longer.
According to Kolmogorov\cite{c1} viscous dissipation occurs only within
the scale of Kolmogorov length. Then, the integral region of dissipation
function (26) which is proportional to the kinematic viscosity is
confined within a small volume of  $O(l_{\kol}^3)$. Dimensional analysis
of (26) gives
\begin{equation}
 \epsilon\sim\frac{\nu}{l^3}\left(\frac{\mvec{q}}{l_{\kol}}\right)^2l_{\kol}^3
\end{equation}
Thus, from (27) through (30) we are given for the order of magnitude of $|\mvec{q}|$ : 
\begin{equation}
 |\mvec{q}|/v\sim R^{7/8}
\end{equation}
It is an indicative of strong in/out mass flow existent (cf. the second
of Eq.(28) of the classical theory) in the vicinity of this extremely
small `energy black hole' where the flow loses kinetic energy $\epsilon$
converted into heat. Inside this region $|\vs|<O(l_{\kol})$ 
the turbulence dies off rapidly towards $q(0)=0$  .There must be, therefore, a drastic variation in the magnitude of $|\mvec{q}|$ which peaks at the boundary ridgeline between dissipation and inertial ranges surrounding the origin ($\vs$=0), and diminishes quickly inside. 

In the neighborhood of this `pseudo'-singularity it is obvious for the following conditions to hold:
\begin{equation}
 \begin{array}{rcl}
  |\mvec{q}|&>> &v \\
  \underbrace{\partial/\partial s_j}_{O(l_{\kol}^{-1})}&>> &\underbrace{\partial/\partial x_j}_{O(l^{-1})} \\
 \end{array}
\end{equation}
The latter condition warrants for local homogeneity in the sense of
Kolmogorov to hold most strictly in the localized eddy space for any
inhomogeneous turbulence. Under these
circumstances, the 6D Navier-Stokes equation (20) together with equation
of continuity reduce to
\begin{equation}
 \partial q_j/\partial s_j=0
\end{equation}
\begin{equation}
 \left(q_r\frac{\partial}{\partial s_r}
  -\nu\frac{\partial^2}{\partial s_r^2}\right)q_j
  +\frac{\partial q_4}{\partial s_j}=0
\end{equation}
which is nothing but the classical (3D) equations for laminar viscous
flows, as disguised through
\begin{equation}
 \vx\rightarrow\vs\;,\;(\vu,p)\rightarrow(\mvec{q},\rho q_4)
\end{equation}
This rule reigns dissipation region as well as the adjacent region of
inertial subrange surrounding it, where   $\mvec{q}$-function diminishes
outward off the pseudo-singularity until flow inhomogeneity starts to
make its appearance. The solution as such, to be pursued in this space,
is `universal', namely, to be valid even for any shear turbulence. 

According to Kolmogorov\cite{c1}, the inertial range is where no viscous
effects are operating. So we may claim that the potential flow is
prevailing there. This speculation is supported by the assertion: `{\it potential flow is the solution of the Navier-Stokes equation in the region where no solid boundary is existent.}'  Now is the case with it, because no physical substances are intervening here in this space.
\section{The pseudo-singularity in the inertial subrange}
The pseudo-singularity, as viewed from the domain of the inertial range, 
looks as if a genuine singularity, whose actual form is now to 
be identified. 

We start with checking if the local isotropy in the classical sense has physical reality. The isotropy assumption is equivalent to 
\begin{equation}
 q_j=s_jQ(s)
\end{equation}
which is the alternative expression of Robertson's theorem\cite{c5}. Substitution of (36) into equation of continuity (33) gives 
\[
\begin{array}{c}
 sdQ/ds+3Q=0 \\
 \yueni\; Q\sim s^{-3}
\end{array}
\]
Obviously, this is a source/ sink flow, corresponding to the velocity
potential 
\begin{equation}
 \phi^{(0)}=\alpha s^{-1}
\end{equation}
If $\alpha$  is positive it represents a sink flow with mass flux
$4\pi\alpha$  vanishing at the origin. Reminding that the
dissipation range $|\vs|\lesssim \;l_{\kol}$ is the `black hole' of turbulent
kinetic energy losing the amount by $\epsilon$ converted into heat every second,
we see that no mass is supposed to be lost. Thus solution (36) for
spherical  isotropy is ruled out. 

Prospective singularity now to take over can be sought within potential
 flow regime as follows: It is obvious that operator
 $\partial ^N/\partial s_1^l\partial s_2^m\partial s_3^n (N=l+m+n)$
 commutes with Laplacean operator, therefore
 \begin{equation}
  \phi^{(N)}=\frac{\partial^N \phi^{(0)}}{\partial s_1^l \partial s_2^m\partial s_3^n}
 \end{equation}
represent a group of potential flows. In particular 
\begin{equation}
 \phi_j^{(1)}=\frac{\partial \phi^{(0)}}{\partial s_j}
\end{equation}
representing a dipole aligned with its axis parallel to $s_j$ direction
is seen to meet the purpose. In fact a dipole is made up of a
pair of sink/source of equal strength, so no mass flux is lost
at this spot. Thus simplest possible candidate for the
pseudo-singularity is {\it axially} isotropic.

The following observation from direct numerical simulation\cite{c6}
helps us draw a more precise picture of our pseudo-singularity: In the
physical space the dissipation occurs only within a confined volume of
`elementary particles' of rod shape, randomly dispersed in turbulent
medium. They all have a shape like a worm with radius $\sigma_0$ of
several Kolmogorov lengths and several ten times of it lengthwise (
$s_0=N\sigma_0$ ), rotating round their own axes. Their lifetime is
about $l_{\kol}/v_{\kol}=(l/v)R^{-1/2}$. So to say, they are like firefly worms
illuminating light for their lifetime as small as of the order of far
subseconds. 

This picture when mapped onto our eddy space is such that two spinning
dipoles having opposite sign and rotation are placed at 
\begin{equation}
 \vs=(s_1,s_2,s_3)=(\pm s_0,0,0)
\end{equation}
respectively, where the axis of symmetry is $s_1$ axis. The direction  j
= 1  is the direction of motion of turbulence-generating body,
representing the only vector prescribing the fluid motion. The flow
field induced by the pair of spinning dipoles
\begin{equation}
 \left.
 \begin{array}{rcl}
  \vq&= &\vq_D+\vq_S \\
  \vq_D&= &\vq_D^+ +\vq_D^- \\
  \vq_S&= &\vq_S^+ +\vq_S^-
 \end{array}\right\}
\end{equation}
with $\vq_D^{\pm}$ and $\vq_S^{\pm}$ standing for velocities induced by dipoles and line vortices, placed at points (40), respectively:
\begin{equation}
 \vq_D^{\pm}=\pm\nabla(\vs)\frac{\partial}{\partial s_1}
  \frac{\alpha}{|\vs\mp\mvec{i}s_0|}
\end{equation}
\begin{equation}
 \vq_S^{\pm}=\pm\frac{\beta}{4\pi}\delta(s_1\mp s_0)
  \frac{\mvec{i}\times(\vs\mp\mvec{i}s_0)}{|\vs\mp\mvec{i}s_0|^3}
\end{equation}
where $\mvec{i}$  is the unit vector designating $s_1$ axis. Expression
(42) is direct consequence of (37) and (39), and that for (43) is
Biot-Savart's law for a line vortex with infinitesimal length directing
$s_1$-axis and with circulation $\beta$. For example, the actual form of
 $q_1(\vs)$ rewritten in axially isotropic form $q_1(s_1,\sigma)$ with
  $\sigma^2=s_2^2+s_3^2$ is,
\begin{equation}
 \begin{array}{rcl}
  q_1(s_1,\sigma)&= &
  \displaystyle
  \frac{\partial^2}{\partial s_1^2}\left\{
  \frac{\alpha}{[(s_1-s_0)^2+\sigma^2]^{1/2}}
 -\frac{\alpha}{[(s_1+s_0)^2+\sigma^2]^{1/2}}
 \right\} \\
  \vspace{0mm}\\
  &= &
  \displaystyle
  \alpha\left\{
  -\frac{2}{[(s_1-s_0)^2+\sigma^2]^{3/2}}
  +\frac{3\sigma^2}{[(s_1-s_0)^2+\sigma^2]^{5/2}}\right.\\
  \vspace{-2mm}\\
  &&\left.
  \displaystyle
     \hspace{5mm}+\frac{2}{[(s_1+s_0)^2+\sigma^2]^{3/2}}
  -\frac{3\sigma^2}{[(s_1+s_0)^2+\sigma^2]^{5/2}}
 \right\} \\
 \end{array}
\end{equation}
Note that there is no contribution from $\vq_S$ to $q_1$.
For $|s_1|>>s_0$ this expression approaches to
\begin{equation}
 q_1\sim
 6\alpha s_0s_1\left(
-\frac{2}{s^5}+\frac{5\sigma^2}{s^7}
     \right)
\end{equation}
which is a quadrupole field, as it should be expected.

Streamlines are shown in Fig.1 of the fictitious flow generated by a
pair of dipoles given by the potential
\begin{equation}
 \left.
 \begin{array}{rcl}
  \phi^{(1)}&= &\partial \phi^{(0)}/\partial s_1 \\
  \phi^{(0)}&= &\alpha \left\{[(s_1-s_0)^2+\sigma^2]^{-1/2}-
   [(s_1+s_0)^2+\sigma^2]^{-1/2} \right\}
 \end{array}\right\}
\end{equation}
This axi-symmetric flow can also be represented using streamfuntion $\psi$
 as
\begin{equation}
 \left.
 \begin{array}{rcl}
  q_{1}&= &\partial \phi^{(1)}/\partial s_1
   = \sigma^{-1} \partial (\sigma \psi)/\partial \sigma\\
  q_{\perp}&= &\partial \phi^{(1)}/\partial \sigma
   = -\sigma^{-1} \partial (\sigma \psi)/\partial s_1
 \end{array}\right\}
\end{equation}
from which we have
\begin{equation}
 \psi=-\sigma \partial \phi^{(0)}/\partial \sigma
\end{equation}
The flow pattern $\psi$ :const. shows quadrupole-like structure at far
field( $|\vs|>> s_0$ ) toward which the longitudinal vortices are
stretched streamwise and then getting thicker. On the returning path to the dipole
core they are chopped off and trim the aspect ratio, getting into the
dissipation region. This picture may serve the qualitative description
of what is actually observed (Fig.2).

Expression (45) gives us estimate for the outer boundary of the locally
homogeneous region. That is also the inner boundary of inhomogeneous region where the first condition of inequality (32) ceases to hold;
\begin{equation}
 |\vq|\sim v
\end{equation}
Let this boundary be defined by $s\sim l_0$, then we have $q\sim \alpha
s_0l_0^{-4}$ from (45), therefore condition (49) is replaced with a more
precise one
\begin{equation}
 v_0\sim \alpha s_0 l_0^{-4}
\end{equation}
where $v_0$ is the characteristic velocity corresponding to $l_0$. They
supplement Kolmogorov formula (27) as
\begin{equation}
 \epsilon\sim v_{K}^3/l_K\sim v_0^3/l_0 \sim v^3/l
\end{equation}
On the opposite side of the inertial subrange $s-s_0 \sim l_{\kol}$,
$\sigma\sim l_{\kol}$, $q$ is estimated from (44) as
\begin{equation}
 q(s_0)\sim \alpha l_{\kol}^{-3},
\end{equation}
We note that the dimensional analysis developed in Sec.4 still holds by
replacing $|\vq|$ with $q(s_0)$ of (52), for instance,
\begin{equation}
 q(s_0)/v\sim R^{7/8}.
\end{equation}
Then, by eliminating $\alpha, q(s_0)$ and $v_0$ from (50) through (53)
we have
\begin{equation}
 l_0/l=(s_0/l)^{4/13}R^{-11/26}
\end{equation}

The relationship between $s_0$ and $l_{\kol}$ or $l$ is yet to be
reconsidered. At present no consensus formula is available for explicit
parameter dependence of worm size $s_0$ .
A recent observation by
numerical experiments\cite{c8} is that $s_0$  is of the order of the
Taylor microscale ($\sim l\:\ol{u'}/v,\; \ol{u'}$ ; r.m.s. of the velocity fluctuation ), according to which the outer boundary of the inertial
range is
\begin{equation}
 l_0/l\sim R^{-11/26}(\ol{u'}/v)^{4/13}
\end{equation}
\section{Power spectrum for inertial subrange}
The actual form of the pseudo-singularity as predicted in the preceding
section enables us to calculate 1D power spectrum for 1D wave number
$k_1$ in the inertial subrange. This spectrum function $P_{11}(k_1)$ is
written, by definition, 
\begin{equation}
 \ol{u_1'^2}=\int_{-\infty}^{\infty}
  P_{11}(k_1)dk_1
\end{equation}
The well-known consequence of the classical dimensional analysis is that
$P_{11}$ , having the dimension of $v^2l\sim\epsilon^{2/3}l^{5/3}$, reads
in the language of wave number space as
\begin{equation}
 P_{11} \sim k_1^{-5/3}
\end{equation}

An alternative look into this formula on our own basis is the following:
The actual form of $P_{11}$ is  obtained through comparing formula (18)
for $\hat{\vx}=\vx$ and $j=l=1$, namely,
\[
 \ol{u_1'^2}=\rp l^3
 \int_{-\infty}^{\infty}dk_1
 \int_{-\infty}^{\infty} dk_2dk_3 f_1f_1^*
\]
with (56), which reads 
\begin{equation}
 P_{11}(k_1)=l^3
  \int_{-\infty}^{\infty}
  dk_2dk_3 f_1f_1^*
\end{equation}
This integral is, upon substitution of (19), transformed into the one in
$\vs$-space as
\begin{equation}
 P_{11}(k_1)=(2\pi l)^{-3}
  \int_{-\infty}^{\infty} ds_1
  \int_{-\infty}^{\infty} d \hat{s}_1 e^{-ik(s_1-\hat{s}_1)}
  \int_{0}^{\infty} \sigma d\sigma q_1(s_1,\sigma)q_1(\hat{s}_1,\sigma)
\end{equation}
Since the integration spans over the whole $\vs$-space, we have yet to
know solution $q_1$ inside the dissipation range [$O(l_K^3)$] where essentially viscous
flow prevails. However, the volume of the dissipation range is by far
the smaller than inertial subrange [$O(l_0^3)$], we may dispense with
potential flow solution (44) to be integrated over its own region. Thus
integral (59) with a small spheroidal regions excepted is to be carried
out. (See Fig. 1.) Then, the integral is shown to be converted into a
double integral as follows 
\begin{equation}
 P_{11}(k_1)=\frac{2}{(2\pi l)^3}
  \int_{0}^{\infty}d(\sigma^2)Q(\sigma^2,k_1)^2
\end{equation}
with $Q(\sigma^2,k_1)$ defined by  
\begin{equation}
 Q(\sigma^2,k_1)=\int_{s_1^\dagger}^{\infty}
  \sin k_1s_1\frac{\partial^2 \phi^{(0)}}{\partial s_1^2}ds_1
 \end{equation}
where boundary contour $s_1=s_1^\dagger(\sigma^2)$ is given by.
\begin{equation}
 s_1^\dagger(\sigma^2)=s_0(1+\delta)\left[
 1-\frac{\sigma^2}{s_0^2(2\delta+\delta^2)}
			      \right]
\end{equation}

The 1D power spectrum calculated is shown in Fig.3. Parameter $\delta$
corresponds  to the slenderness ratio of the worm as
detected by the direct numerical simulation, which is estimated as the
order of $O(10N)^{-1}$ , with $N$ a number of the order of unity to several. For
a certain range of this parameter, the spectrum shows $k_1^{-5/3}$
dependence, then with increase in $\delta$ it transits to $k_1^{-2}$
for $\delta >>1$ , where the pair of dipoles are regarded as a
quadrupole asymptotically.
\section{Conclusions}
6D Navier-Stokes' equation [Eq.(22)] governing turbulence is specialized
to the locally homogeneous dissipation/inertial ranges in the eddy
space. This equation  [Eq.(34)] has revealed existence of a
(pseudo-)singular solution of shape like a volcano with a ridgeline
separating the inertial subrange from the dissipation range inside
which is a energy black hole of turbulence. This finding of anomalous
hike of the velocity wave function ((31) or (53)) in the eddy space has
enabled us to rediscover the universal form of the power spectrum from
the equation of fluid dynamics as distinct from the dimensional analysis
of the classical theory.  In fact, an analytical solution of
one-parameter family defining the ridgeline contour includes a case that
is close to the consensus wave number dependence of -5/3 power law by
Kolmogorov. It is yet left to be answered to eliminate the parameter
dependence. It will be achieved by replacing the inviscid solution
valid only for inertial subrange with a prospective viscous
solution that is uniformly valid throughout dissipation and inertial
regions.
\section{Acknowledgment}
The author wishes to express his thanks to Dr. Shigeru Tachibana for
carrying out numerical calculation for the power spectrum to provide Fig.3.

\newpage
\noindent
{\bf Figure Captions}\\
\\
Figure 1. Pseudo-singularity formed by a pair of dipoles in the inertial
subrange of the eddy space. The shaded part represents the dissipation
region.[See Eq.(63).]\\
Figure 2. Evolution of turbulent vortices visualized by a gigantic
turbulent jet : Volcano eruption of Mt. Saint Helens\cite{c7}.\\
Figure 3. 1D power spectrum $P_{11}$  plotted against 1D wave number $k_1$
with slenderness ratio $\delta$ as the parameter. The calculated
spectrum is averaged over five neighboring points to make the comparison
with -5/3 or -2 power laws easier.
\end{document}